\documentclass[12pt]{article}

\usepackage{texdraw}

\begin{document}

\begin{center}

{\Large \bf Gauge Theories with Cayley-Klein $SO(2;j)$ and $SO(3;{\bf j})$ Gauge Groups} 
\end{center}

\begin{center}
{\large \bf N.A. Gromov} \\
 Department of Mathematics, \\
Komi Science Center UrD RAS \\
Kommunisticheskaya st., 24, Syktyvkar, 167982, Russia \\
E-mail: gromov@dm.komisc.ru
\end{center}

 \begin{center}
{ \bf Abstract}
\end{center}
 
Gauge theories with the orthogonal Cayley-Klein gauge groups $SO(2;j)$ and $SO(3;{\bf j})$ are regarded.
For nilpotent values of the contraction parameters ${\bf j}$  these groups are isomorphic to the non-semisimple
Euclid, Newton, Galilei groups and corresponding matter spaces are fiber spaces with degenerate metrics.
It is shown that the contracted gauge field theories describe the same set of  fields and particle mass as 
$SO(2),\, SO(3)$ gauge  theories, if Lagrangians in the base and  in the fibers all are taken into account. 
Such  theories based on non-semisimple contracted group provide more simple field interactions as compared with the initial ones. 

\section{Introduction}
  
 Gauge field theory was suggested by Yang and Mills \cite{YM-54} and is regarded now as most powerfull method for unified description of fundamental interactions in particle physics, where the compact semisimple Lie groups seem to play the most fundamental roles. For example, in the standard Weinberg-Salam model \cite{W-67}, \cite{S-68} of electroweak theory, the gauge group is $SU(2)\times U(1). $ 
   
  It was realized  by Nappi and Witten \cite{NW-93} that one can also construct gauge theories for some non-semisimple groups (which admit a nondegenerate invariant bilinear form ) and such theories have much simpler structure than the standard theories with semisimple groups.
 Later the gauge theories, $\sigma$-models and solitonic hierarchies for  different non-semisimple groups was investigated \cite{Ts-95}--\cite{V-06}. Contraction of the standard electroweak Weinberg-Salam model to the gauge group $SU(2;\iota)\times U(1)$ was described in   \cite{G-06-2}.
 
In this work we discuss  gauge theories based on non-semisimple orthogonal Cayley-Klein groups which can be obtained from the classical simple groups by contractions. The spaces of the fundamental representations of such  groups are fiber  spaces with degenerate metrics.

\section{Gauge theory  for $SO(2)$ group}
 
   Let $ {\bf R_4}$ is Minkowski space-time: 
    $
    x_\mu x_\mu = x_0^2-x_1^2-x_2^2-x_3^2,  \; \mu=0,1,2,3
    $
  and $\Phi_2$ is the target  space, i.e. the space of fundamental representation of $SO(2)$ group, that elements named matter fields depend on $x \in {\bf R_4}.$
  Gauge transformations:
$\phi'(x)=\omega(\alpha(x))\phi(x), \; \omega(\alpha(x)) \in SO(2)$ or
%$$
\begin{equation}
\left(
\begin{array}{c}
	\phi'_1(x) \\
	\phi'_2(x)
\end{array} \right)
=\left(
\begin{array}{rl}
\cos \alpha(x) & \sin \alpha(x)\\
-\sin \alpha(x) & \cos \alpha(x)	
\end{array}\right) 
\left(
\begin{array}{c}
	\phi_1(x) \\
	\phi_2(x)
\end{array} \right)
\label{1}
\end{equation}
%$$
leave invariant the form  
$\phi^t\phi= \phi_1^2(x)+\phi_2^2(x) $ 
and define Euclid metrics in $\Phi_2.$

   The Lagrangian is written as  \cite{R-99} 
   %$$
   \begin{equation}
   L=-\frac{1}{4}F_{\mu\nu}F_{\mu\nu} + \frac{1}{2}(D_\mu \phi)^{t}D_\mu \phi+ \frac{\mu^2}{2}\phi^{t}\phi -\frac{\lambda}{4}(\phi^{t}\phi)^2, 
\label{2}
\end{equation}   
  % $$
   where covariant derivatives are
   %$$
   \begin{equation}
   D_\mu\phi_1=\partial_\mu\phi_1 + eA_\mu\phi_2, \quad D_\mu\phi_2=\partial_\mu\phi_2 - eA_\mu\phi_1.
   \label{3}
\end{equation}  
   %$$
  Here $e$ is the coupling constant,  $A_\mu(x) $ is the gauge field and the  
    field tensor is defined in the standard way
  $
   F_{\mu\nu}=\partial_\mu A_\nu - \partial_\nu A_\mu. 
   $ 
   
 Higgs mechanism \cite{H-64} is the method of generation mass for gauge fields.
 A Lagrangian  ground state is such configuration of fields $A_\mu, \phi_1, \phi_2, $ that minimize theirs
 energy.
   There are a set of ground states
      %$$
      \begin{equation}
      (\phi_1^{\mbox{vac}})^2 + (\phi_2^{\mbox{vac}})^2 = \phi_0^2, \quad A_\mu^{\mbox{vac}}=\partial_\mu\alpha,\quad \phi_0=\frac{\mu}{\sqrt{\lambda}},
      \label{4}
\end{equation}  
      %$$
 which can be obtained by gauge transformations from
 one of them:
   %$$
   \begin{equation}
   A_\mu^{\mbox{vac}}=0, \quad \phi^{\mbox{vac}}=\left(
\begin{array}{c}
	\phi_0 \\
	0
\end{array} 
\right), \quad \phi_0=\frac{\mu}{\sqrt{\lambda}},     
   \label{5}
\end{equation}  
   %$$   
  as it is shown on Fig. 1.  
  
\begin{figure}[h]
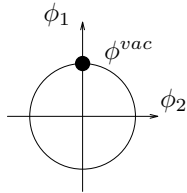
%tb]    
  \centertexdraw{
\drawdim{mm}
\setunitscale 0.5
\bsegment
\move(0 -20)\arrowheadtype t:V \arrowheadsize l:2 w:1 \linewd 0.2 \ravec(0 45)
\move(-20 0)\ravec(40 0)
\textref h:C v:C \htext(24 4) {\footnotesize $\phi_2$}
\textref h:C v:C \htext(-7 27) {\footnotesize $\phi_1$}
\textref h:C v:C \htext(12.0 17) {\footnotesize $\phi^{vac}$}
\move(0 0) \lcir r:14
\move(0 14) \fcir f:0 r:2
%\move(0 0) \rlvec(4 9)
\esegment
}
\caption{ Lagrangian  ground states for $SO(2)$ gauge theory.}
%} 
\end{figure}%[h]%tb]
   
    For small (linear) field exitations   with respect of vacuum %are regarded as follows
$A_\mu (x),\;\phi_1 (x)=\phi_0 +\chi (x) ,\;\phi_2 (x)  $ 
Lagrangian (\ref{2}) can be written as
 %$$
 \begin{equation} 
 L=L^{(2)}+L^{(3)}+L^{(4)},
   \label{6}
\end{equation}  
 %$$ 
    where quadratic in fields $A_\mu, \chi , \phi_2  $ Lagrangian  
   $$
   L^{(2)}=-{\frac{1}{4}B_{\mu\nu}B_{\mu\nu} + \frac{e^2\phi_0^2}{2}B_\mu B_\mu}
   +{\frac{1}{2}(\partial_\mu\chi)^2 -\mu^2\chi^2},
   $$
   %$$
   \begin{equation}
   B_\mu = A_\mu - \frac{1}{e\phi_0}\partial_\mu \phi_2, \quad F_{\mu\nu}=B_{\mu\nu}
   \label{7}
\end{equation}  
   %$$
    describe massive vector field {$B_\mu, \; m_V=e\phi_0=\frac{e\mu}{\sqrt{\lambda}} $} --- 
   gauge field 
      and massive scalar field {$\chi, \; m_\chi=\sqrt{2}\mu $  --- 
    matter field (Higgs boson).
   Field interactions are given  by $L^{(3)}$  and $ L^{(4)}$
   $$
   L^{(3)}=eA_\mu\left(\phi_2\partial_\mu\chi - \chi\partial_\mu\phi_2\right) +\phi_0\chi\left[e^2A_\mu^2 - \lambda\left(\chi^2+\phi_2^2\right)\right],
   $$
   %$$
   \begin{equation}
   L^{(4)}=\frac{1}{2}\left(\chi^2+\phi_2^2\right)\left[e^2A_\mu^2 - \frac{\lambda}{2}\left(\chi^2+\phi_2^2\right)\right],
   \label{8}
\end{equation}  
   %$$
which  include terms of third and fourth order in fields.

\section{Gauge theory for Galilei group. }

\subsection{Galilei group  and Galilei geometry }
  Galilei space  $\Phi_2({\iota})$ and Galilei group $G_2=SO(2;{\iota})$ 
   can be obtained from $\Phi_2 $ and $SO(2)$ by substitution:
   $
   \phi_2 \rightarrow {j}\phi_2, \; \alpha \rightarrow {j}\alpha, 
   $ 
   where contraction parameter takes two values ${j}=1,{\iota}, \;\; {\iota^2}=0,\; \iota/\iota=1. $
  Gauge transformations
%$$
\begin{equation}
\left(
\begin{array}{c}
	\phi'_1(x) \\
	j\phi'_2(x)
\end{array} \right)
=\left(
\begin{array}{rl}
\cos j\alpha(x) & \sin j\alpha(x)\\
-\sin j\alpha(x) & \cos j\alpha(x)	
\end{array}\right) 
\left(
\begin{array}{c}
	\phi_1(x) \\
	j\phi_2(x)
\end{array} \right)
\label{9}
\end{equation}
%$$
leave invariant the form  
$ \phi^t({j})\phi({j})= \phi_1^2+{j^2}\phi_2^2, $ 
which for $ j=1$ define Euclid metrics in $\Phi_2.$
  
For $j={\iota}$  Galilei (degenerate) metrics 
 $
 \phi^t(\iota)\phi(\iota)=\phi_1^2+\iota^2\phi_2^2
 $
in the 2-dim fiber space $\Phi_2(\iota) $
is obtained, where $\{\phi_1\}$ is 1-dim base and $ \{\phi_2\}$ is 1-dim fiber. 
There are {\bf two invariants}: 
$\mbox{inv}_1= \phi_1^2$ under the general transformations
$\phi'({\iota})=\omega({\iota}\alpha)\phi({\iota}), $ where
\begin{equation}   
SO(2;\iota) \ni  \omega({\iota}\alpha)=\left(
\begin{array}{rl}
1 &  {\iota}\alpha \\
- {\iota}\alpha & 1	
\end{array}\right), \; \alpha \in {\bf R}, \quad \omega^t({\iota}\alpha)\omega({\iota}\alpha)=1
\label{10}
\end{equation}
 and $\mbox{inv}_2= \phi_2^2 $ under  transformations in the fiber $(\phi_1=0).$  %: $ \phi_2'=\phi_2.$
   Therefore there are {\bf two metrics}: one in the base and another  in the fiber.

A bundle  of lines through a point on this two  planes has different properties relative to the plane automorphism   \cite{P-65}. 
On  Euclid plane, any two lines of the bundle are transformed to each other by rotation  around the point (see Fig. 2).

\begin{figure}[h]
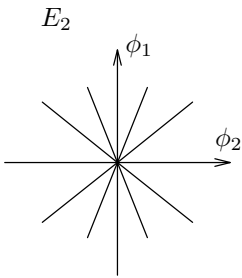
%tb]
\centertexdraw{
\drawdim{mm}
\setunitscale 1.0
%%%  fig E_2 %%%%%%%%%
\move(45 0)
\bsegment
\move(20 5)\arrowheadtype t:V \arrowheadsize l:2 w:1 \linewd 0.2 \ravec(0 30)
\move(5 20)\ravec(30 0)
 \move (20 20) \lvec(10 12)
 \move (20 20) \lvec(30 28)
 \move (20 20) \lvec(10 28)
 \move (20 20) \lvec(30 12)
 \move (20 20) \lvec(24 10)
 \move (20 20) \lvec(24 30)
 \move (20 20) \lvec(16 10)
 \move (20 20) \lvec(16 30)
%\textref h:C v:C \htext(20 0) {\footnotesize a)}
\textref h:C v:C \htext(12 39) {\footnotesize $E_2$}
\textref h:C v:C \htext(23 35.5) {\footnotesize $\phi_1$}
\textref h:C v:C \htext(35 23) {\footnotesize $\phi_2$}
\esegment
%%%  fig 2 G_2 %%%%%%%%%
%\lpatt()
}
\caption{A bundle  of lines on Euclid $E_2$  plane.}
\end{figure}
     
  On  Galilei plane, there is one isolated line   that  do not superposed with any other line of the bundle by Galilei boost (see Fig. 3).
  
\begin{figure}[h]
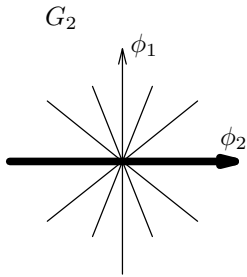
%tb]
\centertexdraw{
\drawdim{mm}
%%%  fig 2 G_2 %%%%%%%%%
\move(45 0)
\bsegment
\move(20 5)\arrowheadtype t:V \arrowheadsize l:2 w:1 \linewd 0.2 \ravec(0 30)
\linewd 1.0
\move(5 20)\ravec(30 0)
\linewd 0.2
 \move (20 20) \lvec(10 12)
 \move (20 20) \lvec(30 28)
 \move (20 20) \lvec(10 28)
 \move (20 20) \lvec(30 12)
 \move (20 20) \lvec(24 10)
 \move (20 20) \lvec(24 30)
 \move (20 20) \lvec(16 10)
 \move (20 20) \lvec(16 30)
%\textref h:C v:C \htext(20 0) {\footnotesize b)}
\textref h:C v:C \htext(12 39) {\footnotesize $G_2$}
\textref h:C v:C \htext(23 35.5) {\footnotesize $\phi_1$}
\textref h:C v:C \htext(35 23) {\footnotesize $\phi_2$}
%\textref h:C v:C \htext(19 20) {      \rule{30mm}{2pt}}  %\color{red}
\esegment
%\lpatt()
}
\caption{A bundle  of lines on Galilei $G_2$   plane.}
\end{figure}

If one interpret these planes in some physical context, then on  Euclid plane all lines must have 
the same physical dimension $[\phi_1]=[\phi_2].$
 On Galilei plane, there are infinite many lines with physical dimension identical with  dimension of the base $[\phi_1]$ and one isolated line in the fiber with some different physical dimension 
 $[\phi_2]\neq[\phi_1]\;$ (see \cite{G-06} for details).
 
\subsection{Gauge theory for Galilei group $G_2.$ }

 Gauge theory for  $SO(2;{\iota})=G_2$   can be obtained from thouse for $SO(2)$  by the substitution 
   %$$
\begin{equation}    
   \phi_1 \rightarrow \phi_1, \quad \phi_2 \rightarrow {j}\phi_2, \quad A_\mu \rightarrow {j}A_\mu, \quad
   F_{\mu\nu} \rightarrow {j}F_{\mu\nu},\quad \mbox{with}\;\; {j=\iota}.
\label{11}
\end{equation}   
   %$$
   %with $ {j=\iota}.$
  Full Lagrangian is splited  on the Lagrangian in the base
   %$$
\begin{equation}   
 { L_b=\frac{1}{2}(\partial_\mu\phi_1)^2+\frac{\mu^2}{2}\phi_1^2  -\frac{\lambda}{4}\phi_1^4,}
\label{12}
\end{equation}     
   %$$
  the  Lagrangian in the fiber $(\approx  {j^2})$
   $$
  {  L_f=-\frac{1}{4}F_{\mu\nu}^2+\frac{1}{2}(\partial_\mu\phi_2)^2+\frac{\mu^2}{2}\phi_2^2+}
   $$
   %$$
\begin{equation}    
 {  +\frac{1}{2}\phi_1^2 (e^2A_\mu^2 - \lambda\phi_2^2) 
   +eA_\mu\left(\phi_2\partial_\mu\phi_1 - \phi_1\partial_\mu\phi_2\right)},   
\label{13}
\end{equation}       
      %$$
      and higher order part $(\approx  {j^4}) $ 
   %$$
\begin{equation}    
  { L_h=\frac{1}{2}\phi_2^2\left( e^2A_\mu^2-\frac{\lambda}{2}\phi_2^2\right)},
\label{14}
\end{equation}    
   %$$
   which disappear for ${j=\iota}.$
 
 %In geometrical interpretation 
 Higgs mechanism  is realized in three steps: 
 
  (i) the Lagrangian  in the base $L_b$ is maximal and the  Lagrangian  in the fiber is equal to zero $L_f=0$ at
  %$$
\begin{equation}  
  \phi_1=\phi_0, \quad \phi_2=0, \quad A_\mu =\frac{1}{e}\partial_\mu\alpha,\quad F_{\mu\nu}=0,\quad \lambda\phi_0^2=\mu^2,
\label{15}
\end{equation}   
  %$$
 where point $M(\phi_0,0) \in \Phi_2({\iota})$ is one of the  ground states;
  
(ii) gauge transformations 
  %$$
\begin{equation}   
  \phi_1'=\phi_1, \quad \phi_2'=\phi_2+\alpha\phi_1, \quad 
  A_\mu' =A_\mu +\frac{1}{e}\partial_\mu\alpha,\quad F_{\mu\nu}'=F_{\mu\nu},
\label{16}
\end{equation}     
  %$$
  applied to $M$ define the set of ground states $\left\{ \phi_1^2=\phi_0^2,\; \phi_2 \in {\bf R}, \;\; 
 A_\mu' =\frac{1}{e}\partial_\mu\alpha \right\}$ as {sphere}  in   $\Phi_2({\iota})$ (see Fig. 4);
  
%\vspace{3mm}
\begin{figure}[h]
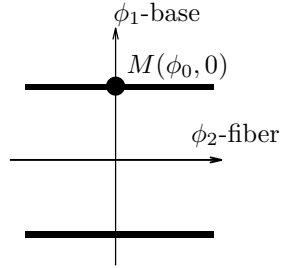
%tb]  
\centertexdraw{
\drawdim{mm}
\setunitscale 0.7
\bsegment
\move(0 -20)\arrowheadtype t:V \arrowheadsize l:2 w:1 \linewd 0.2 \ravec(0 45)
\move(-20 0)\linewd 0.4 \ravec(40 0)
\linewd 0.2
\textref h:C v:C \htext(23 5) {\footnotesize $\phi_2$-fiber}
\textref h:C v:C \htext(8 28) {\footnotesize $\phi_1$-base}
\textref h:C v:C \htext(12 18) {\footnotesize  $M(\phi_0,0)$}
\move(-16 14)\linewd 0.4 \rlvec(32 0)
\move(-16 -14) \rlvec(32 0)
\textref h:C v:C \htext(0 14) { \rule{25mm}{2pt}}
\textref h:C v:C \htext(0 -14) { \rule{25mm}{2pt}}
\move(0 14) \fcir f:0 r:1.8
\esegment
}
\caption{ Ground states for Galilei  gauge theory.}
\end{figure}%[h]%tb] 

  (iii) field excitations arround ground state $M$ are (see Fig. 5)
  %$$
\begin{equation}    
  \phi_1(x)=\phi_0+\chi(x), \quad \phi_2(x),\quad A_\mu(x).
\label{17}
\end{equation}    
  %$$
  
 \begin{figure}[h]
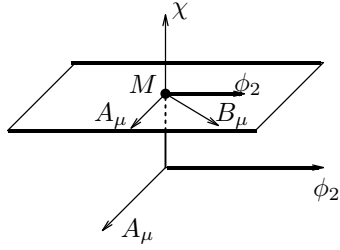
%tb] 
   \centertexdraw{
\drawdim{mm}
\setunitscale 0.7
\bsegment
\move(0 0)\lvec(0 7)
\lpatt(0.5 1) \lvec (0 14)
\lpatt()
\move(0 14)\arrowheadtype t:V \arrowheadsize l:2 w:1 \linewd 0.2 \ravec(0 15)
\move(0 14) \ravec (15 0)
\move(0 14) \ravec (10 -6)
\move(0 14) \ravec (-6.5 -6.5)
\move(0 0) \ravec(30 0)
\move(0 0) \ravec(-12 -12)
\move(-17 20) \lvec(30 20) \lvec(17 7)
\lvec(-30 7) \lvec (-17 20) 
\textref h:C v:C \htext(31 -4) {\footnotesize $\phi_2$}
\textref h:C v:C \htext(-5 -12) {\footnotesize $A_{\mu}$}
\textref h:C v:C \htext(3 30) {\footnotesize $\chi$}
\textref h:C v:C \htext(13 10) {\footnotesize $B_{\mu}$}
\textref h:C v:C \htext(-10 10) {\footnotesize $A_{\mu}$}
\textref h:C v:C \htext(-4 16) {\footnotesize $M$}
 \textref h:C v:C \htext(16 16)  {\footnotesize $\phi_2$}
\textref h:C v:C \htext(6.0 20) {\rule{33mm}{1pt}}
\textref h:C v:C \htext(-6 7) {\rule{33mm}{1pt}}
\textref h:C v:C \htext(7.5 14) {\rule{10mm}{1pt}}
\textref h:C v:C \htext(15 0) {\rule{20mm}{1pt}}

\move(0 14) \fcir f:0 r:1
\esegment
} 
\caption{Field excitations arround ground state $M.$}
\end{figure}     

As a result we obtain 
the  Lagrangian in the base
  $$
  L_b=
 { \frac{1}{2}(\partial_\mu\chi)^2-\mu^2\chi^2 }
  +L_b^{(3)}+ L_b^{(4)},
  $$
   %$$ 
\begin{equation}    
 L_b^{(3)}=  -\lambda\phi_0\chi^3, \quad L_b^{(4)}= -\frac{\lambda}{4}\chi^4,
\label{18}
\end{equation}      
  %$$
  which describe massive scalar  field $\chi, \; m_\chi=\sqrt{2}\mu $ 
  (Higgs boson),  its self-action  $L_b^{(3)},\; L_b^{(4)}$
  and the Lagrangian in the  fiber
   $$
   L_f={-\frac{1}{4}B_{\mu\nu}^2+\frac{e^2\phi_0^2}{2}B_\mu^2} + L_f^{(3)}+L_f^{(4)},
   $$
   $$
  L_f^{(3)}=  eA_\mu\left(\phi_2\partial_\mu\chi - \chi\partial_\mu\phi_2\right) +
  \phi_0\chi \left(e^2A_\mu^2-\lambda\phi_2^2  \right), \quad   
   $$
   %$$
\begin{equation}    
 L_f^{(4)}= \frac{1}{2}\chi^2 (e^2A_\mu^2 -  \frac{\lambda}{2}\phi_2^2),\quad   
   { B_\mu = A_\mu - \frac{1}{e\phi_0}\partial_\mu \phi_2}, 
\label{19}
\end{equation}      
   %$$
   which describe massive vector  gauge  field $B_\mu, \;   m_V=e\phi_0=\frac{e\mu}{\sqrt{\lambda}} $  
   and field interactions.
  
     So in the  theory with Galilei gauge group $G_2$
    matter field (Higgs boson) ${\chi}$ in the base and gauge field ${B_\mu}$ in the fiber have different physical dimensions. Nevertheless, the  mass dimension   of Higgs boson ${m_\chi=\sqrt{2}\mu} $ and
   vector boson ${ m_V=e\phi_0=\frac{e\mu}{\sqrt{\lambda}} }$ are identical and are the same as for
  $SO(2)$ gauge theory.
  More simple field interactions are provided by  Galilei gauge theory as compared to $SO(2)$ one.

\section{Gauge theories for $SO(3;{\bf j})$ groups}

   \subsection{Unified gauge model}

The orthogonal Cayley-Klein group  $SO(3;{\bf j}),\; {\bf j}=(j_1,j_2)$ is defined  \cite{G-90} as the transformation group of the real space $\Phi_3({\bf j}),$ which leave invariant the quadratic form 
\begin{equation}
 \phi^t({\bf j})\phi({\bf j})= \phi_1^2+j_1^2\phi_2^2+j_1^2j_2^2\phi_3^2.  
\label{3-1}
\end{equation}
Here each contraction parameter takes two values $j_k=1,\iota_k,\; k=1,2,\; \iota_k^2=0,\; \iota_1\iota_2=\iota_2\iota_1\neq 0.$
Generators of the Lie algebra $so(3;{\bf j})$ 
\begin{equation}
T_1=\left(
\begin{array}{rrr}
0 & -j_1 & 0\\
j_1 & 0 & 0 \\
0 & 0 &	0
\end{array}\right),\; 
T_2=\left(
\begin{array}{crc}
0 & 0 & -j_1j_2 \\
0 & 0 &	0 \\
j_1j_2 & 0 & 0 
\end{array}\right),\;
T_3=\left(
\begin{array}{rrr}
0 & 0 & 0\\
0 & 0 & -j_2 \\
0 & j_2 &	0
\end{array}\right)
\label{3-2}
\end{equation}
are subject of the commutation relations
\begin{equation}
[T_1,T_2]=j_1^2T_3,\quad [T_2,T_3]=j_2^2T_1,\quad [T_3,T_1]=T_2.
\label{3-3}
\end{equation}
Gauge fields depend on $x\in{\bf R_4}$ and their values are in the algebra $so(3;{\bf j})$
\begin{equation}
A_\mu(x)=gT^aA_\mu^a(x)=g\left(
\begin{array}{ccc}
0 &  -j_1A_\mu^1 & -j_1j_2A_\mu^2 \\
j_1A_\mu^1 & 0 &	-j_2A_\mu^3 \\
j_1j_2A_\mu^2 & j_2A_\mu^3 & 0 
\end{array}\right).
\label{3-4}
\end{equation}
Stress tensor is given by 
\begin{equation}
F_{\mu\nu}(x)=\partial_\mu A_\nu(x)-  \partial_\nu A_\mu(x)+[A_\mu(x),A_\nu(x)]=gT^aF_{\mu\nu}^a(x)
\label{3-5}
\end{equation}
and its components are as follows
$$
F_{\mu\nu}^1=\partial_\mu A_\nu^1-  \partial_\nu A_\mu^1 +j_2^2g(A_\mu^2A_\nu^3-A_\mu^3A_\nu^2),
$$
$$
F_{\mu\nu}^2=\partial_\mu A_\nu^2-  \partial_\nu A_\mu^2 +g(A_\mu^3A_\nu^1-A_\mu^1A_\nu^3),
$$
\begin{equation}
F_{\mu\nu}^3=\partial_\mu A_\nu^3-  \partial_\nu A_\mu^3 +j_1^2g(A_\mu^1A_\nu^2-A_\mu^2A_\nu^1).
\label{3-6}
\end{equation}
Covariant derivative  $D_\mu \phi({\bf j})=[\partial_\mu+gT(A_\mu)]\phi({\bf j})$ 
or in the matrics form 
\begin{equation}
\left(
\begin{array}{r}
D_\mu \phi_1  \\
j_1D_\mu \phi_2  \\
j_1j_2D_\mu \phi_3 
\end{array}\right)=
\left(
\begin{array}{rrr}
\partial_\mu &  -j_1gA_\mu^1 & -j_1j_2gA_\mu^2 \\
j_1gA_\mu^1 & \partial_\mu &	-j_2gA_\mu^3 \\
j_1j_2gA_\mu^2 & j_2gA_\mu^3 & \partial_\mu 
\end{array}\right)
\left(
\begin{array}{r}
 \phi_1  \\
j_1 \phi_2  \\
j_1j_2\phi_3 
\end{array}\right)
\label{3-7}
\end{equation}
acts on the field components  $\phi({\bf j}) $ as 
$$
D_\mu \phi_1=\partial_\mu \phi_1 -gj_1^2(A_\mu^1\phi_2 + j_2^2A_\mu^2\phi_3),\quad
%$$
%$$
D_\mu \phi_2=\partial_\mu \phi_2+g(A_\mu^1\phi_1 - j_2^2A_\mu^3\phi_3),
$$
\begin{equation}
D_\mu \phi_3=\partial_\mu \phi_3+g(A_\mu^2\phi_1 + A_\mu^3\phi_2).
\label{3-8}
\end{equation}

The complete Lagrangian $L({\bf j})=L_A({\bf j})+L_\phi({\bf j})$ of the model is defined as the sum of the gauge fields
Lagrangian 
\begin{equation}
L_A({\bf j})=\frac{1}{8g^2}\mbox{Tr}(F_{\mu\nu}({\bf j}))^2= -\frac{1}{4}[j_1^2(F_{\mu\nu}^1)^2+j_1^2j_2^2(F_{\mu\nu}^2)^2+j_2^2(F_{\mu\nu}^3)^2]
\label{3-9}
\end{equation}
and the matter fields Lagrangian 
\begin{equation}
L_\phi({\bf j})=\frac{1}{2}(D_\mu \phi({\bf j}))^{t}(D_\mu \phi({\bf j})) - V(\phi;{\bf j}),
\label{3-10}
\end{equation}
where potential is taken in the standard form 
\begin{equation}
V(\phi;{\bf j})=\left(\frac{\sqrt{\lambda}}{2}\phi^{t}({\bf j})\phi({\bf j}) - \frac{\mu^2}{2\sqrt{\lambda}}  \right)^2.
\label{3-11}
\end{equation}
In an explicit form the gauge fields Lagrangian is given by
\begin{equation}
L_A({\bf j})= -[j_1^2({\cal F}_{\mu\nu}^1)^2+j_1^2j_2^2({\cal F}_{\mu\nu}^2)^2+ 
j_2^2({\cal F}_{\mu\nu}^3)^2]- 
j_1^2j_2^2[L_A^{(3)}({\bf j})+L_A^{(4)}({\bf j})].
\label{3-12}
\end{equation}
The terms of the third and fourth order in fields are equal to
$$
L_A^{(3)}({\bf j})=2g[{\cal F}_{\mu\nu}^1(A_\mu^2A_\nu^3-A_\mu^3A_\nu^2) + {\cal F}_{\mu\nu}^2(A_\mu^3A_\nu^1-A_\mu^1A_\nu^3)+
{\cal F}_{\mu\nu}^3)(A_\mu^1A_\nu^2-A_\mu^2A_\nu^3)],
$$
\begin{equation}
L_A^{(4)}({\bf j})=g^2[j_2^2(A_\mu^2A_\nu^3-A_\mu^3A_\nu^2)^2+(A_\mu^3A_\nu^1-A_\mu^1A_\nu^3)^2 +
j_1^2(A_\mu^1A_\nu^2-A_\mu^2A_\nu^3)^2],
\label{3-13}
\end{equation}
where ${\cal F}_{\mu\nu}^a=\partial_\mu A_\nu^a-  \partial_\nu A_\mu^a$ is  stress tensor in a flat space.

Ground states of $L({\bf j})=L_A({\bf j})+L_\phi({\bf j})$ are such  fields configuration, that  $L_A({\bf j})=0 $ 
and potential (\ref{3-11}) is minimal.      
The ground states are realized by the fields 
$$
A_\mu(x;{\bf j})=\omega(x;{\bf j})\partial_\mu \omega^{-1}(x;{\bf j}),\;
\; \omega(x;{\bf j})\in SO(3;{\bf j}),
$$
\begin{equation}
\phi^t({\bf j})\phi({\bf j})= \phi_1^2+j_1^2\phi_2^2+j_1^2j_2^2\phi_3^2=\phi_0^2, \; \phi_0=\frac{\mu}{\sqrt{\lambda}}
\label{3-14}
\end{equation}
which for  $A_\mu(x;{\bf j})=0 $ belong to the sphere in the matter field space of the radius  $\phi_0 $
and can be obtained by the gauge transformations
$$
\phi (x;{\bf j})=\omega(x;{\bf j})\phi^{vac},\; 
$$
\begin{equation}
A'_\mu(x;{\bf j})= \omega(x;{\bf j})A_\mu(x;{\bf j})\omega^{-1}(x;{\bf j}) + \omega(x;{\bf j})\partial_\mu \omega^{-1}(x;{\bf j})
\label{3-15}
\end{equation}
applied to one of the ground states $M.$ For  simplicity this ground state  $M$ is taken in the form 
\begin{equation}
A_\mu(x;{\bf j})=0,\;  (\phi^{vac})^t=(\phi_0,0,0)^t.
\label{3-16}
\end{equation}

Next step in Higgs mechanism  is to introduce small field excitations arround the ground state $M$
\begin{equation}
A_\mu^a(x),\quad  \phi_1(x)=\phi_0+\chi(x),\;\;  \phi_2(x),\;\;  \phi_3(x)
\label{3-17}
\end{equation} 
and to write down the complete Lagrangian for the new fields
\begin{equation}  
{\cal L}({\bf j})={\cal L}^{(2)}({\bf j})+{\cal L}^{(3)}({\bf j})+{\cal L}^{(4)}({\bf j}).
\label{3-18}
\end{equation} 
The second order in fields Lagrangian 
$$
{\cal L}^{(2)}({\bf j})=\frac{1}{2}(\partial_\mu\chi)^2 -\mu^2\chi^2 +
j_1^2\left[-\frac{1}{4}(B^1_{\mu\nu})^2 + \frac{g^2\phi_0^2}{2}(B^1_\mu)^2 \right]+
$$
\begin{equation}  
+j_2^2\left[-\frac{1}{4}({\cal F}^3_{\mu\nu})^2  \right]+
j_1^2j_2^2\left[-\frac{1}{4}(B^2_{\mu\nu})^2 + \frac{g^2\phi_0^2}{2}(B^2_\mu)^2 \right],
\label{3-19}
\end{equation} 
where the new fields are introduced 
\begin{equation}
   B^1_\mu = A^1_\mu + \frac{1}{g\phi_0}\partial_\mu \phi_2, \quad 
   B^2_\mu = A^2_\mu + \frac{1}{g\phi_0}\partial_\mu \phi_3,
   \label{3-20}
\end{equation}
describe one massive scalar matter field   {$\chi, \; m_\chi=\sqrt{2}\mu \;$  (Higgs boson), 
two massive vector fields $B^k_\mu, \;k=1,2$ with the identical masses
$ m_V=g\phi_0=\frac{g\mu}{\sqrt{\lambda}} $
and one massless field   $A^3_\mu. $
Field interactions are described by the terms of the third  ${\cal L}^{(3)}({\bf j})$ 
 and fourth ${\cal L}^{(4)}({\bf j})$ order in fields 
   $$
   {\cal L}^{(3)}({\bf j})=  -\lambda\phi_0\chi^3 +  j_1^2\left\{ - \lambda\phi_0\chi(\phi_2^2+j_2^2\phi_3^2) + g\left[A_\mu^1\left(\chi\partial_\mu\phi_2- \phi_2\partial_\mu\chi\right)\right.\right.
   $$
   $$
\left. +j_2^2 A_\mu^2\left(\chi\partial_\mu\phi_3- \phi_3\partial_\mu\chi\right)+
j_2^2 A_\mu^3\left(\phi_2\partial_\mu\phi_3- \phi_3\partial_\mu\phi_2\right) \right]+
   $$
   $$
 \left.+g^2\phi_0 \left[ A_\mu^1\left(A_\mu^1\chi- j_2^2A_\mu^3\phi_3 \right)+ 
 j_2^2A_\mu^2\left(A_\mu^2\chi + A_\mu^3\phi_2 \right)   \right] \right\} -
   $$
   $$
-j_1^2j_2^2\frac{g}{2}\left[{\cal F}^1_{\mu\nu}\left(A_\mu^2A_\nu^3- A_\mu^3A_\nu^2  \right)+
{\cal F}^2_{\mu\nu}\left(A_\mu^3A_\nu^1- A_\mu^1A_\nu^3  \right)+ 
{\cal F}^3_{\mu\nu}\left(A_\mu^1A_\nu^2- A_\mu^2A_\nu^1  \right) \right],   
   $$
   $$
{\cal L}^{(4)}({\bf j})=-\frac{\lambda}{4}\chi^4 +j_1^2\frac{1}{2}\left\{-\lambda\chi^2\left(\phi_2^2+j_2^2\phi_3^2 \right) -j_1^2\frac{\lambda}{2}\left(\phi_2^2+j_2^2\phi_3^2 \right)^2 \right. +
   $$
   $$
+g^2\left[j_1^2\left(A_\mu^1\phi_2+j_2^2A_\mu^2\phi_3 \right)^2+
 \left(A_\mu^1\chi - j_2^2A_\mu^3\phi_3 \right)^2 + j_2^2\left(A_\mu^2\chi+A_\mu^3\phi_2 \right)^2\right]-  
   $$
\begin{equation}   
-j_2^2g^2\left.\left[j_2^2\left(A_\mu^2A_\nu^3 - A_\mu^3A_\nu^2 \right)^2 + 
 \left(A_\mu^3A_\nu^1 - A_\mu^1A_\nu^3 \right)^2 + j_1^2\left(A_\mu^1A_\nu^2 - A_\mu^2A_\nu^1 \right)^2 \right]   \right\}.
\label{3-21}
\end{equation}     

It is worth to note that gauge theories with Cayley-Klein gauge group $SO(3;{\bf j})$ can be obtained from $SO(3)$
gauge theory by the substitution
   $$ 
\chi \rightarrow \chi,\; \phi_2 \rightarrow j_1\phi_2,\; \phi_3 \rightarrow j_1j_2\phi_3,\; 
   $$
\begin{equation} 
 A^1_\mu \rightarrow j_1A^1_\mu,\quad A^2_\mu \rightarrow j_1j_2A^2_\mu,\quad A^3_\mu \rightarrow j_2A^3_\mu,\; 
\label{3-22}
\end{equation}   
or 
\begin{equation} 
 B^1_\mu \rightarrow j_1B^1_\mu,\quad B^2_\mu \rightarrow j_1j_2B^2_\mu,\quad 
 {\cal F}^3_{\mu\nu} \rightarrow j_2{\cal F}^3_{\mu\nu}
\label{3-22-1}
\end{equation}  
for the new fields.

\subsection{Gauge model for  Euclid group  $E_3$}  %=SO(3;\iota_1,j_2)$}

For nilpotent value of the first parameter $j_1=\iota_1$ rotation group $SO(3)$ is contracted to the non-semisimple group $SO(3;\iota_1,j_2),$ which is isomorphic to the Euclid group $E_3.$ The matter space metrics is degenerated
$\phi^t(\iota_1)\phi(\iota_1)= \phi_1^2+\iota_1^2(\phi_2^2+j_2^2\phi_3^2) $
and $\Phi_3(\iota_1)$ becomes the fiber space with one dimensional base $\{\phi_1\} $ and two dimensional fiber
  $\{\phi_2,\phi_3\}. $ Accordingly two Lagrangians are appeared: first in the base
\begin{equation}
{\cal L}_b(\iota_1)=\frac{1}{2}(\partial_\mu\chi)^2 -\mu^2\chi^2 -\lambda\phi_0\chi^3 -\frac{\lambda}{4}\chi^4
-j_2^2\frac{1}{4}({\cal F}^3_{\mu\nu})^2, 
\label{3-23}
\end{equation}
which describe Higgs boson  $\chi $ with the standard mass $m_\chi=\sqrt{2}\mu, $ its self-action   and the massless field $A^3_\mu;$ second Lagrangian  $(\approx j_1^2)$ in the fiber
   $$
{\cal L}_f(\iota_1)= -\frac{1}{4}(B^1_{\mu\nu})^2 + \frac{g^2\phi_0^2}{2}(B^1_\mu)^2 
+ j_2^2\left[-\frac{1}{4}(B^2_{\mu\nu})^2 + \frac{g^2\phi_0^2}{2}(B^2_\mu)^2 \right]+
   $$
\begin{equation}   
+ {\cal L}_f^{(3)}(\iota_1)+ {\cal L}_f^{(4)}(\iota_1),   
\label{3-24}
\end{equation}  
which describe two massive fields $B^1_\mu,\; B^2_\mu $ with the identical masses $m_V=g\phi_0.$
Field interactions in the fiber are given by the terms  $(\approx j_1^2)$ of the third
 ${\cal L}_f^{(3)}(\iota_1) $ and forth  ${\cal L}_f^{(4)}(\iota_1) $ order in fields, where
  $$
{\cal L}_f^{(4)}(\iota_1)=\frac{1}{2}\left\{-\lambda\chi^2\left(\phi_2^2+j_2^2\phi_3^2 \right) \right.
+g^2\left[ \left(A_\mu^1\chi - j_2^2A_\mu^3\phi_3 \right)^2 + j_2^2\left(A_\mu^2\chi+A_\mu^3\phi_2 \right)^2\right]-  
   $$
 \begin{equation}   
-j_2^2g^2\left.\left[j_2^2\left(A_\mu^2A_\nu^3 - A_\mu^3A_\nu^2 \right)^2 + 
 \left(A_\mu^3A_\nu^1 - A_\mu^1A_\nu^3 \right)^2 +  \right]   \right\}
\end{equation} 
and ${\cal L}_f^{(3)}(\iota_1) $ is given by (\ref{3-21}) without the term  $-\lambda\phi_0\chi^3 $ and total multiplier
$j_1^2.$ 
The proportional to $ j_1^4$  forth order terms are equal to zero, i.e. field interactions are more simple as compared with $SO(3)$ model, whereas the fields themselves and their masses are the same as for $SO(3)$ gauge model.

\subsection{Gauge model for  Newton group  $N_3$}

For nilpotent value of the second parameter $j_2=\iota_2$ rotation group $SO(3)$ is contracted to the non-semisimple group $SO(3;j_1,\iota_2),$ which is isomorphic to the Newton group $N_3.$ The matter space metrics is degenerated
$\phi^t(\iota_2)\phi(\iota_2)= \phi_1^2+j_1^2\phi_2^2+\iota_2^2j_1^2\phi_3^2 $
and $\Phi_3(\iota_2)$ becomes the fiber space with two dimensional base $\{\phi_1,\phi_2\} $  and one dimensional fiber
  $\{\phi_3\}. $  Lagrangian in the base 
\begin{equation}
{\cal L}_b(\iota_2)=\frac{1}{2}(\partial_\mu\chi)^2 -\mu^2\chi^2 -\frac{1}{4}(B^1_{\mu\nu})^2 + \frac{g^2\phi_0^2}{2}(B^1_\mu)^2 + {\cal L}_b^{(3)}(\iota_2)  + {\cal L}_b^{(4)}(\iota_2) 
\label{3-26}
\end{equation} 
describe Higgs boson  $\chi, $ massive boson $B^1_\mu$ and interactions in the form
  $$
{\cal L}_b^{(3)}(\iota_2)=-\lambda\phi_0\chi \left(\chi^2+j_1^2\phi_2^2 \right) 
+j_1^2\left[gA_\mu^1\left(\chi\partial_\mu\phi_2 -\phi_2\partial_\mu\chi \right)  \right],  
  $$
\begin{equation}
  {\cal L}_b^{(3)}(\iota_2)= -\frac{\lambda}{4}\chi^4 +
j_1^2\frac{1}{2}\left[  \left(\chi^2+j_1^2\phi_2^2 \right)\left(g^2(A_\mu^1)^2-\lambda\phi_2^2 \right) +
 j_1^2\frac{\lambda}{2}\phi_2^4  \right].
\label{3-27}
\end{equation} 

Lagrangian in the fiber takes the form
\begin{equation}
{\cal L}_f(\iota_2)=-\frac{1}{4}({\cal F}^3_{\mu\nu})^2  +
j_1^2\left[-\frac{1}{4}(B^2_{\mu\nu})^2 + \frac{g^2\phi_0^2}{2}(B^2_\mu)^2 \right] + {\cal L}_f^{(3)}(\iota_2)
 + {\cal L}_f^{(4)}(\iota_2) ,
\label{3-27-1}
\end{equation} 
where
   $$
{\cal L}_f^{(3)}(\iota_2)= j_1^2\left\{-\lambda\phi_0\chi\phi_3^2 +  \right.
   $$
   $$
+g\left[A_\mu^2\left(\chi\partial_\mu\phi_3- \phi_3\partial_\mu\chi\right)+
A_\mu^3\left(\phi_2\partial_\mu\phi_3- \phi_3\partial_\mu\phi_2\right) \right]+ 
 %  $$
 %  $$
 g^2\phi_0 A_\mu^2\left(A_\mu^2\chi + A_\mu^3\phi_2 \right)- 
   $$
   $$
\left. - \frac{g}{2}\left[{\cal F}^1_{\mu\nu}\left(A_\mu^2A_\nu^3- A_\mu^3A_\nu^2  \right)+
{\cal F}^2_{\mu\nu}\left(A_\mu^3A_\nu^1- A_\mu^1A_\nu^3  \right)+ 
{\cal F}^3_{\mu\nu}\left(A_\mu^1A_\nu^2- A_\mu^2A_\nu^1  \right) \right]\right\}, 
   $$
   $$
{\cal L}_f^{(4)}(\iota_2)=j_1^2\frac{1}{2}\left\{-\lambda\phi_3^2\left(\chi^2+j_1^2\phi_2^2 \right) +\right.
   $$
   $$
+g^2\left[ \left(A_\mu^2\chi +A_\mu^3\phi_2  \right)^2 + 2j_1^2A_\mu^1A_\mu^2\phi_2\phi_3 -2A_\mu^1A_\mu^3\chi\phi_3    \right]- 
   $$
\begin{equation}
 \left. - g^2\left[\left(A_\mu^3A_\nu^1 -A_\mu^1A_\nu^3 \right)^2 +j_1^2 \left(A_\mu^1A_\nu^2 -A_\mu^2A_\nu^1 \right)^2         \right]  \right\}.   
\label{3-27-2}
\end{equation} 
So Lagrangian in the fiber describe the massless field $A_\mu^3,$  massive boson $ B_\mu^2 $ and field interactions.

\subsection{Gauge model for  Galilei group  $G_3$}

For $j_1=\iota_1,\; j_2=\iota_2$ the group $SO(3;\iota) $ is isomorphic to Galilei group  $G_3.$
The space  $\Phi_3(\iota_1)$ becomes twice fiber space, since two dimensional fiber $\{\phi_2,\phi_3\} $ in its part is
the fiber space with the base $\{\phi_2\}$ and the fiber $\{\phi_3\}.$ There are three invariants:
 $\mbox{inv}_1=\phi_1^2,\;\;\mbox{inv}_2=\phi_2^2,\;\;\mbox{inv}_3=\phi_3^2. $
Therefore we have Lagrangian in the base 
\begin{equation}
{\cal L}_b(\iota)=\frac{1}{2}(\partial_\mu\chi)^2 -\mu^2\chi^2 -\lambda\phi_0\chi^3 -\frac{\lambda}{4}\chi^4, 
\label{3-28}
\end{equation}  
 Lagrangian in the first fiber $(\approx j_1^2)$  
 \begin{equation}
{\cal L}_{f_1}(\iota)=
-\frac{1}{4}(B^1_{\mu\nu})^2 + \frac{g^2\phi_0^2}{2}(B^1_\mu)^2  + {\cal L}_{f_1}^{(3)}(\iota)
 + {\cal L}_{f_1}^{(4)}(\iota),
\label{3-29}
\end{equation} 
where
   $$
{\cal L}_{f_1}^{(3)}(\iota)=\phi_0\chi\left[g^2\left(A_\mu^1 \right)^2 -\lambda\phi_2^2\right]+
gA_\mu^1\left(\chi\partial_\mu\phi_2 - \phi_2\partial_\mu \right),   
   $$
\begin{equation}
{\cal L}_{f_1}^{(4)}(\iota)=\chi^2\left[g^2\left(A_\mu^1 \right)^2 -\frac{\lambda}{2}\phi_2^2\right],
\label{3-30}
\end{equation} 
 Lagrangian in the second fiber  $(\approx j_1^2j_2^2)$
\begin{equation}
{\cal L}_{f_2}(\iota)=
-\frac{1}{4}(B^2_{\mu\nu})^2 + \frac{g^2\phi_0^2}{2}(B^2_\mu)^2  + {\cal L}_{f_2}^{(3)}(\iota)
 + {\cal L}_{f_2}^{(4)}(\iota),
\label{3-31}
\end{equation} 
where
   $$
{\cal L}_{f_2}^{(3)}(\iota)=g^2\phi_0 A_\mu^2\left(A_\mu^2\chi + A_\mu^3\phi_2 \right)+
   $$
   $$
+g\left[A_\mu^2\left(\chi\partial_\mu\phi_3- \phi_3\partial_\mu\chi\right)+
A_\mu^3\left(\phi_2\partial_\mu\phi_3- \phi_3\partial_\mu\phi_2\right) \right], 
   $$
\begin{equation}
{\cal L}_{f_2}^{(4)}(\iota)=\frac{1}{2}\left\{-\lambda\chi^2 \phi_3^2+
+g^2 \left(A_\mu^2\chi +A_\mu^3\phi_2  \right)^2 
 - g^2\left(A_\mu^3A_\nu^1 -A_\mu^1A_\nu^3 \right)^2   \right\}.  
\label{3-32}
\end{equation}
 
For the two parameter contraction, unlike one parameter ones, the fibering in the gauge field space do not 
coincide with  the fibering in the matter field space, therefore in gauge field Lagrangian (\ref{3-9})
there is the term $(\approx j_2^2),$ which it is necessary to regard as one more Lagrangian
\begin{equation}
{\cal L}_{g}(\iota)=-\frac{1}{4}\left({\cal F}^3_{\mu\nu} \right)^2.
\label{3-33}
\end{equation} 

Thus, for completely contracted Galilei group $G_3$ each field is described by own Lagrangian: Higgs boson --- by Lagrangian in the base ${\cal L}_{b}(\iota),$  massive bosons --- by Lagrangian in the first ${\cal L}_{f_1}(\iota)$
and in the second ${\cal L}_{f_2}(\iota) $ fibers of the matter field space  $\Phi_3(\iota),$ massless field  $A^3_\mu$ 
--- by Lagrangian ${\cal L}_{g}(\iota) $ in the fiber of the gauge field space, which do not coincides with the fibers of $\Phi_3(\iota).$
Nevertheless one can speak about unified description of all fields because all fields are generated by one gauge group.

\section{Conclusion}

 The contracted Cayley-Klein group is  the motion group  of its   fundamental representation space,
 which has degenerate metrics and some set of invariants with respect of this group \cite{G-90}.
 This means that in gauge theories with contracted gauge groups the matter field spaces and the gauge field spaces 
 are the fiber spaces.  For the  complete description of a physical system in such space  it is necessary to regard  the complete set of Lagrangians: in the  base and  in the all fibers \cite{G-06}.
Only in this case the gauge field theory with degenerate metrics in target (matter) field space describe  the same set of  fields and particle mass as  non-contracted one.

 Although in the case of degenerate metrics  there are several Lagrangians nevertheless one can speak  about unified description of all fields because all fields are generated by the one contracted non-semisimple gauge group. In other words, unified description of gauge fields means one gauge group rather than one Lagrangian.

     Since it is the structure constants that determine the interactions and since under contractions of Lie group some  structure constants of its algebra turn to zero, the gauge field theory based on non-semisimple  contracted group  provide  more simple field interactions as compared with initial one.  

This work was partially supported by Russian Foundation for Basic Research under  grant 07-01-00374.

\end{document}